\documentclass[showkeys,floats,floatfix]{revtex4}
\usepackage{graphicx}
\usepackage{gensymb}
\usepackage{amsmath}
\usepackage[version=3]{mhchem}
\usepackage{upgreek}

\begin{document}

\title{Charge State Hysteresis in Semiconductor Quantum Dots}

%\email[These authors contributed equally to this work]{}
\author{C. H. Yang}
\author{A. Rossi}
\email[Electronic mail: ]{a.rossi@unsw.edu.au}
\author{N. S. Lai}
\author{R. Leon}
\author{W. H. Lim}
\author{A. S. Dzurak}
\affiliation{Australian Research Council Centre of Excellence for Quantum Computation and Communication Technology and School of Electrical Engineering \& Telecommunications, The University of New South Wales, Sydney 2052, Australia}

\date{\today}

\begin{abstract}
Semiconductor quantum dots provide a two-dimensional analogy for real atoms and show promise for the implementation of scalable quantum computers. Here, we investigate the charge configurations in a silicon metal-oxide-semiconductor double quantum dot tunnel coupled to a single reservoir of electrons. By operating the system in the few-electron regime, the stability diagram shows hysteretic tunnelling events that depend on the history of the dots charge occupancy. We present a model which accounts for the observed hysteretic behaviour by extending the established description for transport in double dots coupled to two reservoirs. We demonstrate that this type of device operates like a single-electron memory latch.
\end{abstract}

%\pacs{}

\keywords{quantum dots, silicon, nanoelectronics, quantum computing.}% during submission APL

\maketitle
\begin{figure}[]
\includegraphics[scale=0.75]{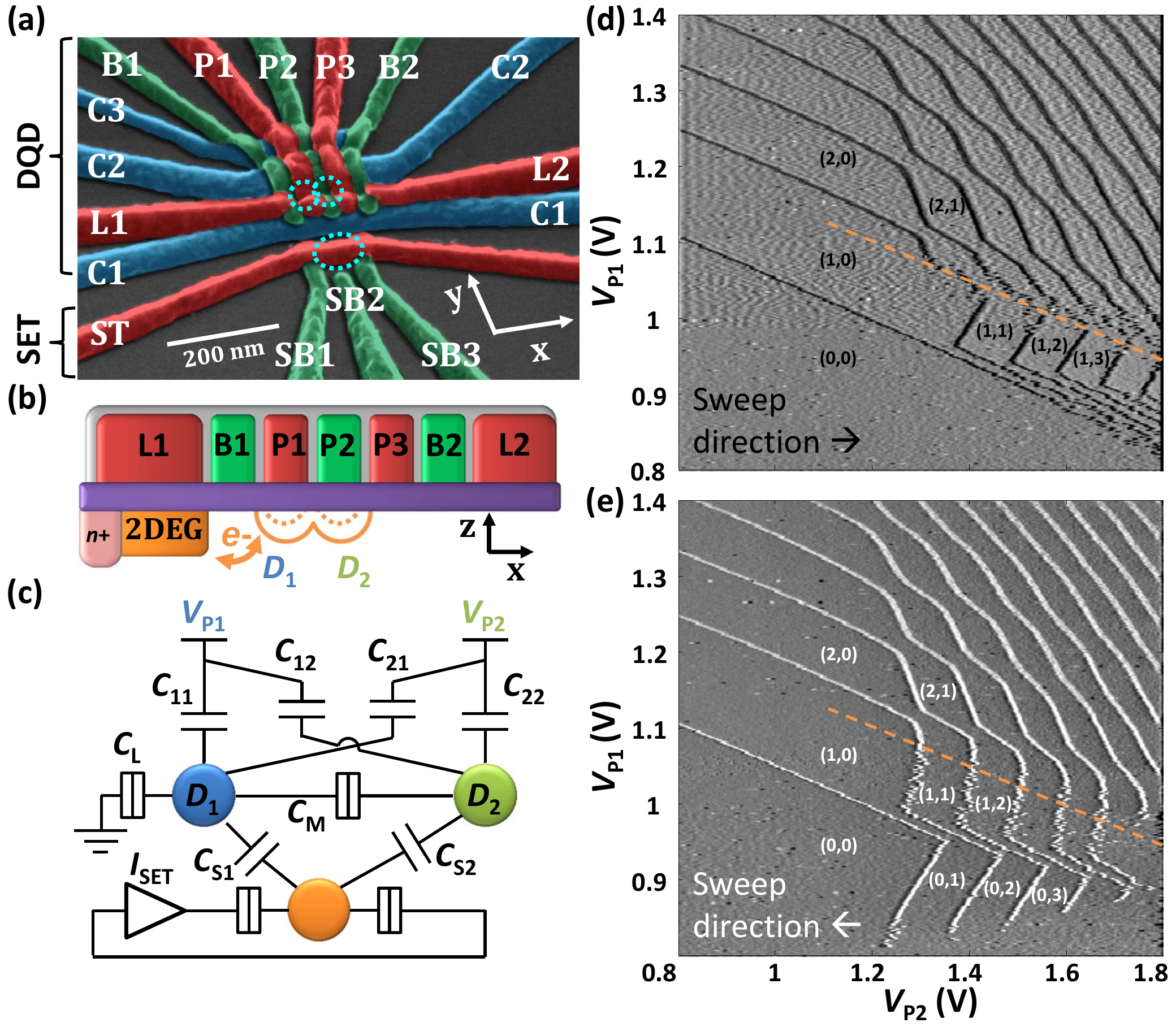}
\caption{(a) False-colour SEM micrograph of a device similar to the one studied. Dashed circles indicate the regions where the DQD and the SET island are defined. (b) Schematic cross-sectional view of the MOS architecture. Metallic gates are indicated in red and blue, the insulating Al$_y$O$_x$ layer is in grey and the SiO$_2$ layer is shown in purple. Dotted conduction band profile illustrates a double dot configuration, while solid profile corresponds to a single merged dot. (c) Schematic circuit of the single-lead DQD system capacitively coupled to the SET electrometer. (d) Experimental stability diagram measured by sensing individual charge transitions (black lines) with the SET detector and dynamical feedback technique. $V_\textup{P2}$ is swept by increasing the voltage, and $V_\textup{L1}=1.6$~V, $V_\textup{C1}=-0.1$~V, $V_\textup{C2}=-0.4$~V, $V_\textup{C3}=2.0$~V, $V_\textup{B1}=1.12$~V, $V_\textup{P3}=V_\textup{B2}=V_\textup{L2}=$~GND. The region with hysteretic transitions lies below the dashed line. (e) Similar diagram as in (d) but for decreasing $V_\textup{P2}$ voltage sweeps. }
\label{dev}
\end{figure}
\begin{figure}[]
\includegraphics[scale=0.7]{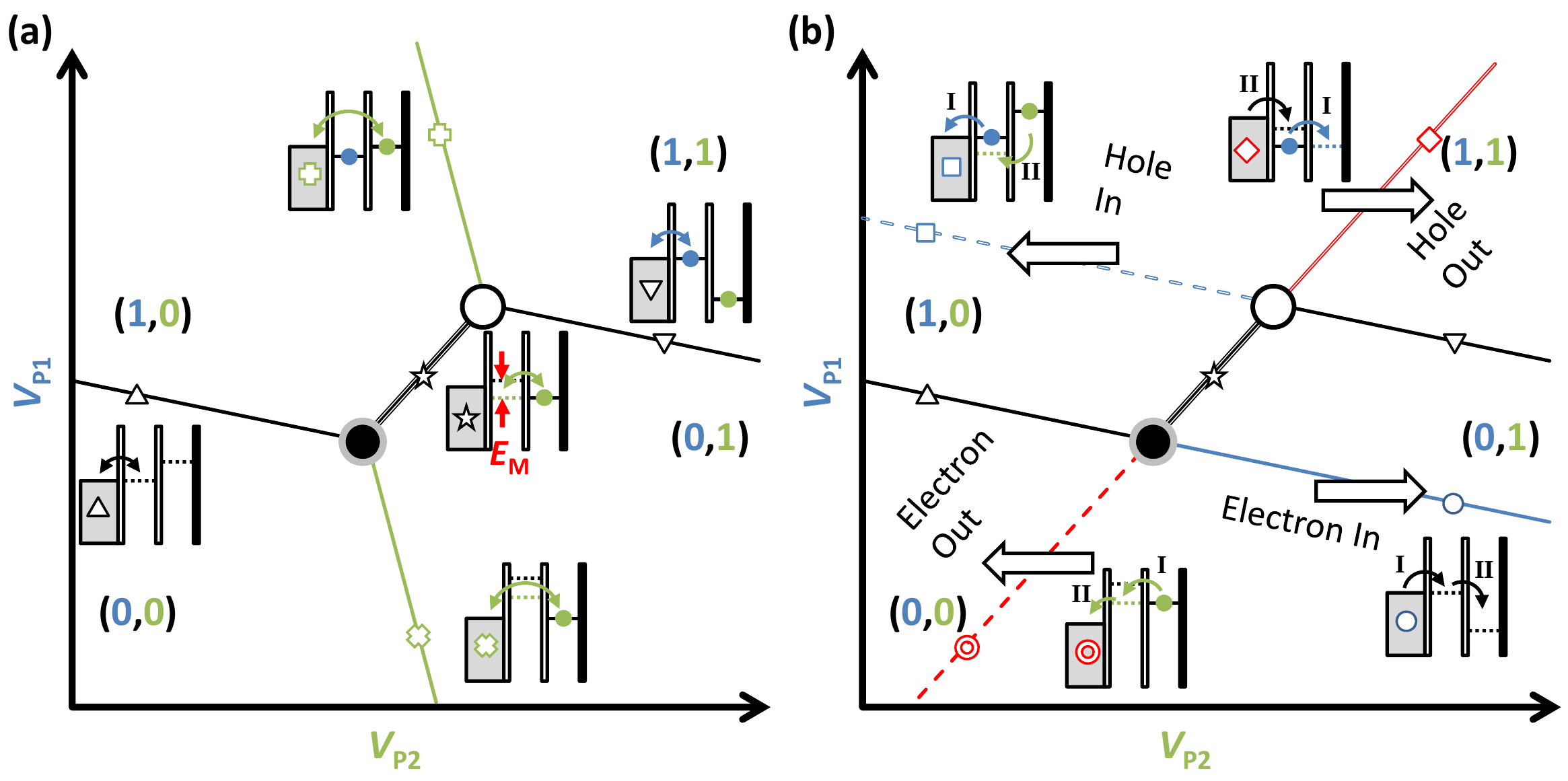}
\caption{Schematic unit cells of charge stability diagrams for the single-lead DQD system (a) in the presence of cotunnelling-mediated transitions, and (b) in the case co-tunnelling is inhibited and hysteresis occurs. Cotunnelling-mediated events are depicted as solid green lines. Hysteretic charge transitions are accompanied by conditional arrows. An arrow pointing rightwards (leftwards) indicates a transition taking place for increasing (decreasing) electron occupancy in D2% by increasing (decreasing) $V_\textup{P2}$ 
. Triple points for electron (hole) transfer processes are indicated by full (empty circles). The schematic energy diagrams show the configurations of the system electrostatic potentials for the corresponding symbol-coded transition line. The order of sequential tunnelling events is indicated.}
\label{trans}
\end{figure}
\begin{figure}[]
\includegraphics[scale=0.75]{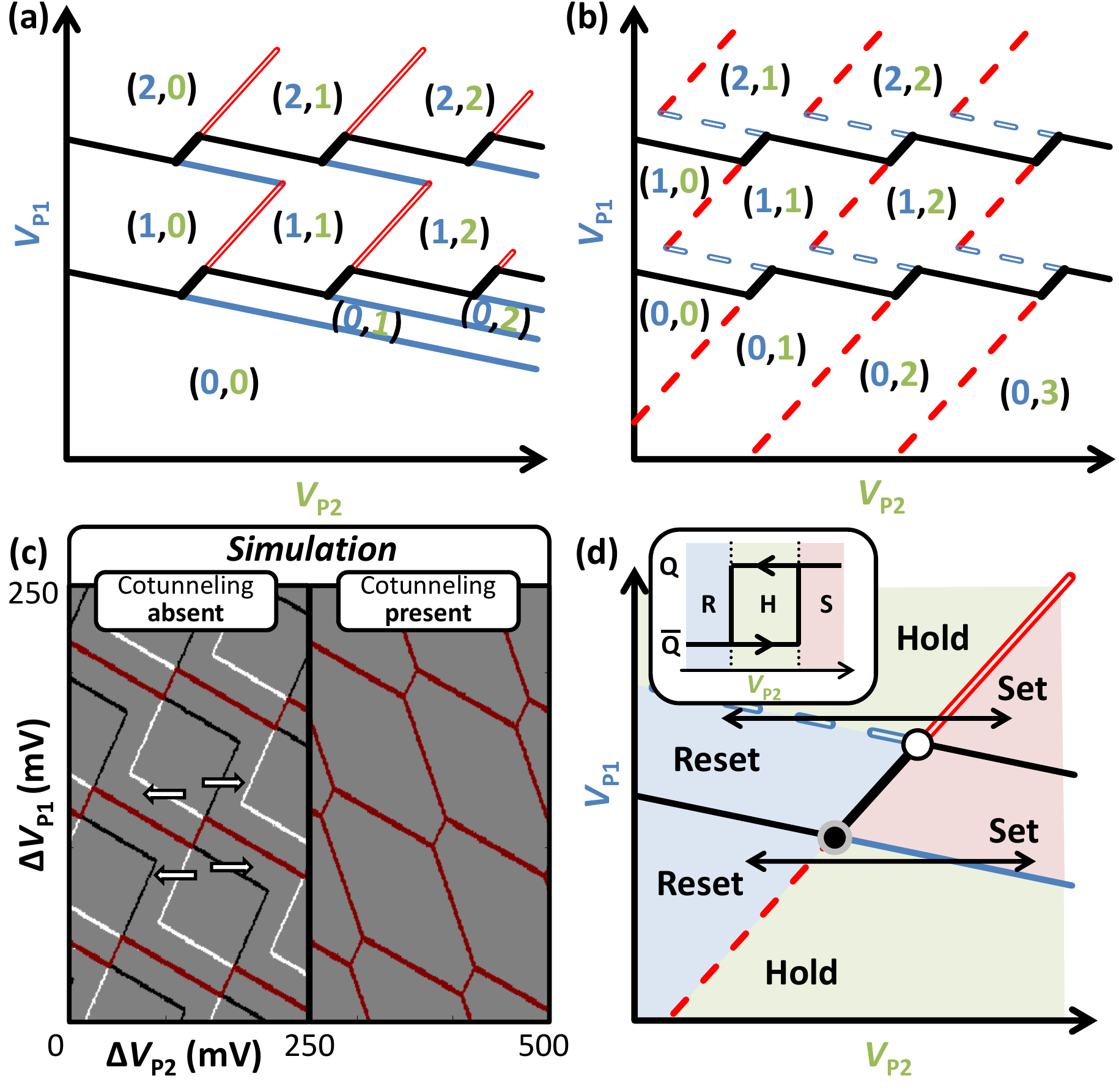}
\caption{(a) Schematic charge stability diagram of a single-reservoir DQD system for increasing number of electrons in D2. The relevant hysteretic transition lines are shown in blue and red. (b) Similar plot as in (a) but for decreasing occupancy in D2. (c) Comparison between simulated charge stability diagrams with co-tunnelling events inhibited (left-hand side) and allowed (right-hand side). Non-hysteretic transtions are indicated in red, while the hysteretic ones are shown in black and white with the relevant conditional arrows. Capacitances used for the simulations are extracted from experiments: $C_{11}=1.98$~aF, $C_{12}=0.51$~aF, $C_{21}=0.29$~aF, $C_{22}=1.45$~aF, $C_\textup{M}=4.0$~aF. (d) Correspondence between the DQD hysteretic stability map and the memory latch operations. Inset: S/R latch transitions from logic state Q to $\overline{Q}$ and vice versa,  implemented at fixed $V_\textup{P1}$ voltage (horizontal two-headed arrows in the main panel).}
\label{sim}
\end{figure}

In the last three decades lithographically defined semiconductor quantum dots have been the focus of extensive research efforts~\cite{kou,wiel,floris}, and have attracted large interest for a number of applications, such as solid-state quantum computing~\cite{loss}, quantum dot cellular automata~\cite{orlov,mitic,perez}, quantum electrical metrology~\cite{peko}, as well as cryogenic temperature measurement~\cite{temp1} and regulation~\cite{temp2}. The electrical properties of these systems are typically investigated either via electron transport between two-dimensional electron gas (2DEG) reservoirs tunnel coupled to the dots~\cite{hanson_rev} or by detecting charge and spin states with on-chip electrometers~\cite{elze}. In the context of quantum computing, the use of charge sensing has become the method of choice to perform non-invasive measurements of the coherent quantum bit states that exist in the dots~\cite{petta,hrl}. Remote detection has, therefore, made it unnecessary to have an electrical current flow and has led to the realization of systems in which only one lead or none is used~\cite{elze,hanson,nishi,simm,our_natcom,ourprb,dimiprb,gorman}. While the absence of 2DEG reservoirs can be beneficial to both suppress thermal fluctuations induced by electrical noise~\cite{yurke,myapl2} and conveniently scale up these systems, hysteretic behaviour is observed that may complicate the tuning of charge states~\cite{myapl1,henryAIP}. Such hysteresis can, however, be exploited for the implementation of single-electron memory devices~\cite{stone1,stone2}.\\\indent Here, we investigate a silicon double quantum dot (DQD) that is tunnel-coupled to a single 2DEG reservoir and capacitively coupled to a single-electron transistor (SET) used to detect individual charge transitions in the DQD. When the system is operated in the few-electron regime the stability diagram reveals hysteresis in the DQD occupancy, with the occurrence of charge transitions depending upon the history of the charge states. We present a model that accounts for the characteristic features observed by extending the conventional description of electron transport through double-lead DQDs~\cite{wiel}. We also show that this system functions as a single-electron Set/Reset (S/R) memory latch.\\\indent
Our devices are metal-oxide-semiconductor (MOS) planar structures fabricated on a high-purity, near-intrinsic natural silicon substrate. Three layers of Al/Al$_y$O$_x$ gates are defined via electron-beam lithography and deposited on a 8-nm-thick SiO$_2$ gate oxide~\cite{angus,lim}. A 2DEG accumulation layer is locally induced at the Si/SiO$_2$ interface upon application of positive gate voltages. Tunable tunnel barriers are selectively formed in the 2DEG by reducing the bias of individual gate electrodes. Figure ~\ref{dev}(a) shows a scanning electron micrograph (SEM) image of a device similar to the one used for the experiments. Two dots are formed under gates P1 and P2. Planar electrostatic confinement is achieved by negatively biasing gates C1 and C2, while gates P3, B2 and L2 are kept at fixed ground potential. Gate C3 is used to induce a 2DEG that acts as a screen to mitigate parasitic effects. A 2DEG reservoir is induced under gate L1 which extends to a heavily n-type doped region acting as an ohmic contact. This reservoir is tunnel-coupled to the DQD by a tunnel barrier formed under gate B1. The inter-dot tunnel barrier is due to the oxidized aluminium layer present between gates P1 and P2, as illustrated in Fig.~\ref{dev}(b). As a result, by modifying the voltage applied to these gates, both the inter-dot coupling and the occupancy of each dot are affected. The remaining gates, are biased in a similar manner to define and control an SET, capacitively coupled to the DQD and used as an electrometer.\\\indent 
The experiments are performed in a dilution refrigerator with an estimated SET electron temperature of approximately 300 mK. A lock-in amplifier is used to measure the current signal in the SET upon modification of the DQD charge states via the relevant control gates [see Fig.~\ref{dev}(c)]. In order to maximize the electrometer sensitivity, we use a dynamical feedback technique~\cite{henryAIP} so that the detector's operating point is virtually unaffected by slow drift or sudden rearrangement of charge.\\\indent Figures~\ref{dev}(d,e) show the charge stability diagrams of the DQD obtained for $V_\textup{P2}$ voltage scans in opposite directions. The diagrams can be divided into three main regions. For high voltages on both P1 and P2 (upper right corners in both plots) we observe straight parallel features which are the signature of charge transitions in a single-dot system. This is consistent with a largely transparent inter-dot barrier due to the high positive voltages applied to the control gates. For lower voltages (central region of both diagrams) the well-known honeycomb pattern associated with a DQD system appears~\cite{wiel}. This is clear indication that a DQD is formed once the control gate voltages are reduced to make the inter-dot barrier sufficiently opaque. Finally, by further lowering the voltages (bottom region for both diagrams) the few-electron regime is entered, and the two stability maps show evident discrepancies. As discussed next, this can be explained by the fact that, in a single-reservoir DQD system, the equilibrium charge configurations depend on the electron occupancy history of the dots.\\\indent
In Fig.~\ref{trans} the unit cells from charge stability plots for different inter-dot tunnelling regimes are schematically illustrated. The black solid lines represent charge transitions relevant to a change of occupancy in D1. Since this dot is directly tunnel coupled to the electron reservoir, these transitions have the same characteristics as those in a conventional honeycomb diagram for double-lead DQDs and, hence, they are not hysteretic. By contrast, transitions highlighted in blue and red in Fig.~\ref{trans}(b) are hysteretic and occur when the occupancy in D2 changes. Unlike the case of a two-lead DQD system, \textit{here the electron number in D2 can only be modified via tunnelling through D1, so that once an electron has tunnelled into D2, it is trapped there unless the control voltage is sufficiently reversed}. Therefore, the observed hysteresis is a consequence of Coulomb blockade in D1 combined with a significant mutual inter-dot electrostatic coupling ($E_\textup{M}$). Specifically, the presence of an extra electron in D2 shifts the potential of D1 by an amount $E_\textup{M}\approx 2.4$~meV causing a voltage shift in the location of these transitions. This explains the dependence of the transition positions on the occupancy of D2 and, ultimately, on the direction of the voltage sweep, as the directional arrows in Fig.~\ref{trans}(b) indicate.\\\indent As Fig.~\ref{trans}(b) shows, in the stability plot of a single reservoir DQD, each triple point corresponds to the intersection of two hysteretic and two non-hysteretic boundaries of equilibrium charge configurations. As for double-lead DQD, these triple points represent the loci of three degenerate charge state configurations~\cite{wiel}. 
This description allows one to sketch stability diagrams that fully account for the hysteretic phenomena. In Fig.~\ref{sim}(a) the plot relevant for an increasing number of electrons in D2 is shown. In this case, the hysteretic transitions indicated in Fig.~\ref{trans}(b) with solid lines are relevant, in combination with all the non-hysteretic events. The boundaries of different equilibrium charge configurations are then seen to be consistent with those observed in the measurements for increasing $V_\textup{P2}$ in Fig.~\ref{dev}(d). Similarly, for decreasing occupancy in D2, hysteretic transitions indicated by dashed lines in Fig.~\ref{trans}(b) are combined with the non-hysteretic ones, as sketched in Fig.~\ref{sim}(b). This is consistent with the experimental data for decreasing $V_\textup{P2}$ voltage in Fig.~\ref{dev}(e).\\\indent
It is noteworthy that the stability diagrams of Fig.~\ref{dev}(d) and (e) do not show any hysteretic behaviour for larger voltages on the two control gates, i.e. above the dashed lines in Fig.~\ref{dev}(d,e). This is consistent with the presence of cotunnelling-mediated charge transitions. By increasing the gate voltages, the inter-dot tunnel barrier becomes increasingly transparent, as confirmed by the observed shift from DQD to single quantum dot behaviour. The enhanced inter-dot tunnel coupling then makes co-tunnelling processes more likely to occur~\cite{wiel,geer}. In this situation, electrons in dot D2 can access the 2DEG reservoir by co-tunnelling through D1, even though D1 is in a Coulomb blockaded state~\cite{geer}. As a consequence, dot D2 becomes virtually tunnel coupled to the reservoir and a hysteresis-free stability map, typical of double-lead DQD, is obtained. \\\indent
In order to corroborate our interpretation, we performed Monte-Carlo simulations~\cite{simon} of the charge stability  characteristic for the measured single-lead DQD. The system is modelled as a network of tunnelling resistors and capacitors, as well as gate coupling capacitors [see Fig.~\ref{dev}(c)] whose numerical values are extracted from the experiments. Figure~\ref{sim}(c) shows the simulated stability plots in which co-tunnelling events are inhibited (left-hand side) or allowed (right-hand side) via algorithm selection. When co-tunnelling is absent, the simulated stability diagram contains all the hysteretic charge transitions at once. The reason for this lies in the nature of the Monte-Carlo approach. This method considers all possible tunnelling events from randomly generated input voltages and charge configurations, and evaluates individual event probabilities as an aggregate over several simulation runs. We note that the regions of fixed charge configuration are consistent with those obtained from the measurements in Fig.~\ref{dev}(d,e) and the cartoon of Fig.~\ref{sim} (a,b). When co-tunnelling is accounted for, the simulated stability diagram shows the conventional honeycomb pattern consistent with the measurements for high inter-dot coupling. This confirms that a \textit{virtual} double lead DQD system arises in the presence of co-tunnelling.\\\indent
We finally note that this device can be operated as a single-electron S/R memory latch, as schematically illustrated in Fig.~\ref{sim}(d). By sweeping the relevant control gate voltages across hysteretic transitions, one can implement \textit{set}/\textit{reset} operations, whereas by scanning the area between hysteretic features a \textit{hold previous state} operation will result. \\\indent 
In conclusion, we have operated a single-lead silicon DQD down to the few-electron regime. The fact that one of the two dots is not directly tunnel coupled to an electron reservoir results in hysteretic charge transitions in the stability diagram. These can be attributed to the effect of strong inter-dot electrostatic coupling in combination with Coulomb blockade. We observe that the hysteretic behaviour is suppressed by the occurrence of co-tunnelling events.
\\\indent This work was financially supported by the Australian Research Council (CE110001027), the Australian Government, and the U.S. Army Research Office (Contract No W911NF-13-1-0024), and by the Australian National Fabrication Facility for device fabrication. The authors acknowledge useful discussions with C. M. Cheng and thank Dave Barber for technical support.
\bibliography{ref_hyst}

%\begin{thebibliography}{15}

%\bibitem{Angus}
%S. J. Angus, A. J. Ferguson, A. S. Dzurak and R. G. Clark, NanoLetters \textbf{7}, 2051 (2007).

%\end{thebibliography} 

\end{document}